\documentclass[twocolumn]{aastex63}

\usepackage{longtable}
\usepackage{amsmath}
\received{July 23, 2022}
\revised{Sept. 18, 2022}
\accepted{Sept. 19, 2022}
\published{Oct. 13, 2022}
\shorttitle{JWST SL analysis of SMACS0723}
\shortauthors{Pascale et al.}
\begin{document}

\title{Unscrambling the lensed galaxies in JWST images behind SMACS 0723}

%% The \author command is the same as before except it now takes an optional
%% argument which is the 16 digit ORCID. The syntax is:
%% \author[xxxx-xxxx-xxxx-xxxx]{Author Name}
%%

\author[0000-0002-2282-8795]{Massimo Pascale}
\affiliation{Department of Astronomy, University of California, 501 Campbell Hall \#3411, Berkeley, CA 94720, USA}

\author[0000-0003-1625-8009]{Brenda L.~Frye}
\affiliation{Department of Astronomy/Steward Observatory, University of Arizona, 933 N. Cherry Avenue, Tucson, AZ 85721, USA}

\author[0000-0001-9065-3926]{Jose Diego}
\affiliation{FCA, Instituto de Fisica de Cantabria (UC-CSIC), Av.  de Los Castros s/n, E-39005 Santander, Spain}

\author[0000-0001-6278-032X]{Lukas J. Furtak}
\affiliation{Physics Department, Ben-Gurion University of the Negev, P. O. Box 653, Be’er-Sheva, 8410501, Israel}

\author[0000-0002-0350-4488]{Adi Zitrin}
\affiliation{Physics Department, Ben-Gurion University of the Negev, P. O. Box 653, Be’er-Sheva, 8410501, Israel}

\author[0000-0002-8785-8979]{Tom Broadhurst}
\affiliation{Department of Physics, University of the Basque Country UPV/EHU, E-48080 Bilbao, Spain}
\affiliation{DIPC, Basque Country UPV/EHU, E-48080 San Sebastian, Spain}
\affiliation{Ikerbasque, Basque Foundation for Science, E-48011 Bilbao, Spain}

\author[0000-0003-1949-7638]{Christopher J. Conselice} % conselice@gmail.com
\affiliation{Jodrell Bank Centre for Astrophysics, University of Manchester, Oxford Road, Manchester, M13\,9PL, U.K.} % [

\author[0000-0003-2091-8946]{Liang Dai}
\affiliation{Department of Physics, University of California, 366 Physics North MC 7300, Berkeley, CA. 94720, USA}

\author[0000-0002-8919-079X]{Leonardo Ferreira} % conselice@gmail.com
\affiliation{Centre for Astronomy and Particle Physics, University of Nottingham,
Nottingham, UK} % [

\author[0000-0003-4875-6272]{Nathan J. Adams} % conselice@gmail.com
\affiliation{Jodrell Bank Centre for Astrophysics, University of Manchester, Oxford Road, Manchester, M13\,9PL, U.K.} % [

\author[0000-0001-9394-6732]{Patrick Kamieneski} 
\affiliation{Department of Astronomy, University of Massachusetts, Amherst, MA 01003, USA} % [

\author[0000-0002-7460-8460]{Nicholas Foo}
\affiliation{Department of Astronomy/Steward Observatory, University of Arizona, 933 N. Cherry Avenue, Tucson, AZ 85721, USA}

\author[0000-0003-3142-997X]{Patrick Kelly}
\affiliation{School of Physics and Astronomy, University of Minnesota, 116 Church Street SE, Minneapolis, MN 55455, USA}

\author[0000-0003-1060-0723]{Wenlei Chen}
\affiliation{School of Physics and Astronomy, University of Minnesota, 116 Church Street SE, Minneapolis, MN 55455, USA}

\author[0000-0003-4220-2404]{Jeremy Lim}
\affiliation{Department of Physics, The University of Hong Kong, Pokfulam Road, Hong Kong}

\author[0000-0002-7876-4321]{Ashish K. Meena}
\affiliation{Physics Department, Ben-Gurion University of the Negev, P. O. Box 653, Be’er-Sheva, 8410501, Israel}

\author[0000-0003-3903-6935]{Stephen M.~Wilkins} %
\affiliation{Astronomy Centre, University of Sussex, Falmer, Brighton BN1 9QH, UK}
\affiliation{Institute of Space Sciences and Astronomy, University of Malta, Msida MSD 2080, Malta}

\author[0000-0003-0883-2226]{Rachana Bhatawdekar} %
\affiliation{European Space Agency, ESA/ESTEC, Keplerlaan 1, 2201 AZ Noordwijk, NL}

\correspondingauthor{Massimo Pascale}
\email{massimopascale@berkeley.edu}

\author[0000-0001-8156-6281]{Rogier A.~Windhorst} % Rogier.Windhorst@gmail.com
\affiliation{School of Earth \& Space Exploration, Arizona State University, Tempe, AZ\,85287-1404, USA}
\affiliation{Department of Physics, Arizona State University, Tempe, AZ\,85287-1504, USA}

%\author[0000-0001-9411-3484]{Miriam Golubchik}
%\author[0000-0001-6278-032X]{Lukas J. Furtak}
%\author[0000-0002-7876-4321]{Ashish K. Meena}
%\author[0000-0002-0350-4488]{Adi Zitrin}
%\author[0000-0002-0350-4488]{Someone et al.}
%\affiliation{Physics Department,
%Ben-Gurion University of the Negev, P.O. Box 653,
%Be'er-Sheva 84105, Israel}



\begin{abstract}

The first deep field images from the \textit{James Webb Space Telescope} (JWST) of the galaxy cluster SMACS~J0723.3-7327 reveal a wealth of new lensed images at uncharted infrared wavelengths, with unprecedented depth and resolution. Here we securely identify 14 new sets of multiply imaged galaxies totalling 42 images, adding to the five sets of bright and multiply-imaged galaxies already known from Hubble Space Telescope data. 
We find examples of arcs crossing critical curves, allowing detailed community follow-up, such as JWST spectroscopy for precise redshift determinations, and measurements of the chemical abundances and of the detailed internal gas dynamics of very distant, young galaxies. One such arc contains a pair of compact knots that are magnified by a factor of hundreds, and features a microlensed transient. We also detect an Einstein cross candidate only visible thanks to JWST's superb resolution.  Our \emph{parametric} lens model is available through the following link\footnote{\url{https://www.dropbox.com/sh/gwup2lvks0jsqe5/AAC2RRSKce0aX-lIFCc9vhBXa?dl=0}}, and will be regularly updated using additional spectroscopic redshifts. The model is constrained by 16 of these sets of multiply imaged galaxies, three of which have spectroscopic redshifts, and reproduces the multiple images to better than an rms of $0.5^{\prime \prime}$, allowing for accurate magnification estimates of high-redshift galaxies. The intracluster light extends beyond the cluster members, exhibiting large-scale features that suggest a significant past dynamical disturbance.   This work represents a first taste of the enhanced power JWST will have for lensing-related science. 
\end{abstract}



\section{Introduction}\label{sec:intro}
%%%%%%%%%%%%%%%%%%%%%%%%%%%%%%%%%%%%%%%%%%%%%%%%%%%

%Gravitational lensing has supplied 

At long last, the first deep images from the {\it James Webb Space Telescope} (JWST) were delivered on July 11, 2022.
The target is the central region of SMACS~J0723.3-732 (SMACS0723 hereafter; $z_d$\,=\,0.39), which is part of the southern extension of the MACS sample \citep{Ebeling2010FinalMACS,ReppEbeling2018}.  SMACS0723 is a massive strong-lensing (SL) galaxy cluster which is rich in galaxy images distorted by the gravitational lensing effect, as was seen in recent \emph{Hubble Space Telescope} (HST) imaging taken as part of the \textit{Reionization Lensing Cluster Survey} \citep[RELICS;][]{Coe2019RELICS}. The $Planck$-derived mass is high 8.39\,$\times$\,10$^{14}$\,M$_{\odot}$, and there were nine galaxies at $z>5.5$ and two candidates at $z=7$ based on their photometric redshift estimation \citep{Salmon2020HighzRelics,strait21}. Three far-infrared sources have also been detected in the field using the {\it Herschel Space Observatory} \citep{Sun2022}. 

A lens model for this cluster based on these previous HST images was published recently \citep{Golubchik2022}. This model uses the \texttt{Light-Traces-Mass} (\texttt{LTM}) method, which assumes that the galaxy light traces the underlying stellar and dark matter (DM) but does not assume a parametrized mass function for the DM \citep[][]{Broadhurst2005a,Zitrin2009_cl0024,Zitrin2014CLASH25}. 
In \cite{Golubchik2022}, the authors identify five strongly lensed galaxies and derive the spectroscopic redshifts for three of those systems using publicly-available ESO \textit{Very Large Telescope} (VLT) \textit{Multi Unit Spectroscopic Explorer} \citep[MUSE;][]{Bacon2010MUSE} data. 
Two additional models are also available, one based on \texttt{Lenstool} \citep{Jullo2007Lenstool,Sharon2022} and the other on \texttt{Glafic} \citep{Oguri2012SL}. These models, however, do not use the spectroscopic information from MUSE.
\\

\begin{figure*}
    \centering
    \includegraphics[width=18.2cm, trim=0.5cm 0cm 0cm 0cm]{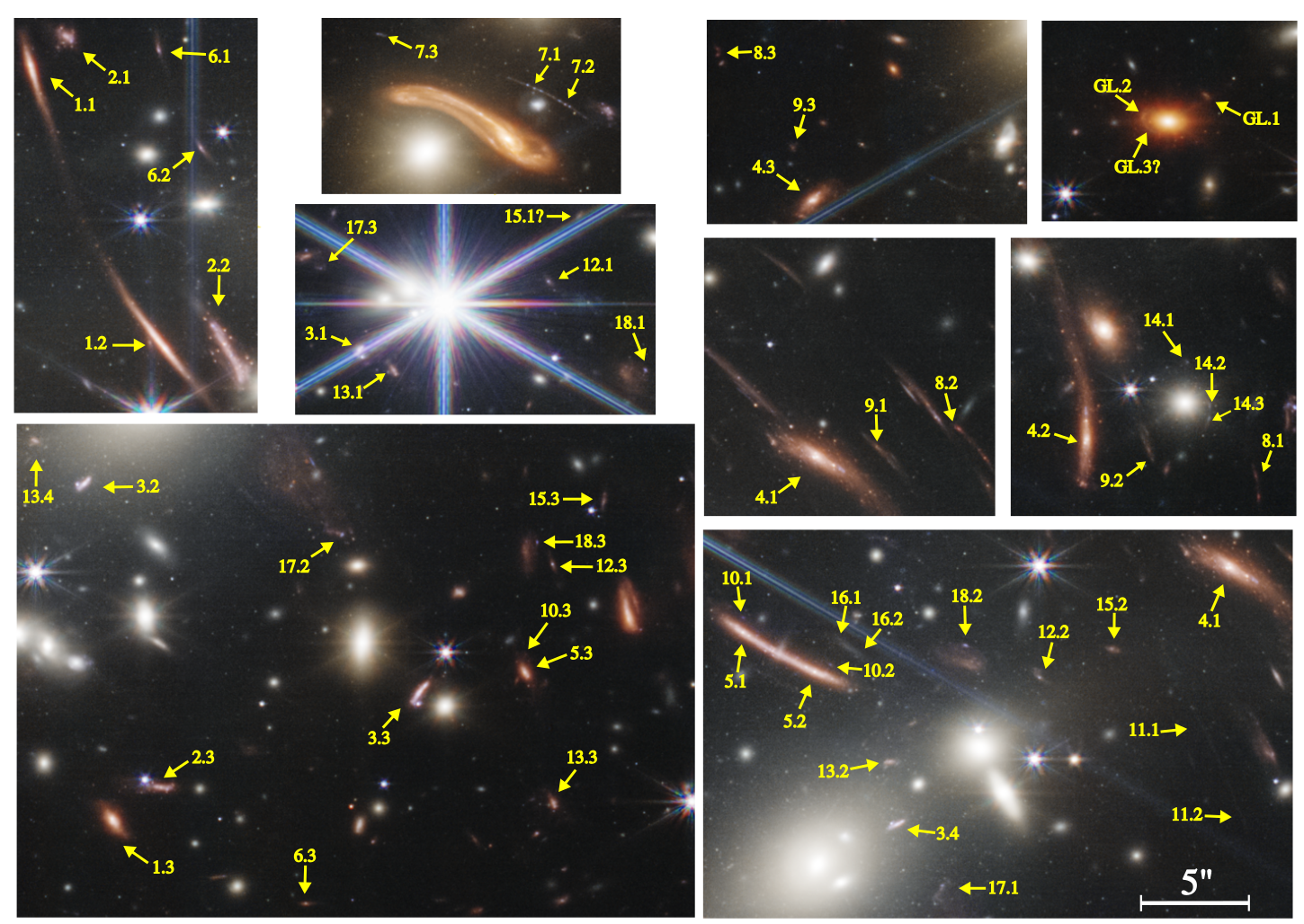}
    \caption{Image stamps depicting the 19 image multiplicities in the SMACS0723 field, as labeled. The publicly-available color image of SMACS0723 is depicted, which is valuable for the identification of image family members by their similar colors and morphological components.  We also require that the model be able to reproduce the positions of the members of each image family.  A 5$^{\prime\prime}$ bar is shown for reference.}
    \label{fig:imstamps}
\end{figure*}

This first set of images of SMACS0723 from JWST represents a milestone for not only astronomy, but science in general. As with its predecessor, HST, this first image from JWST reveals the distant universe with incredible detail. Some of the most prominent arcs detected with the HST RELICS program now show multiple unresolved substructures in JWST previously unseen (see for instance system 4 in Figure~\ref{fig:imstamps}). These substructures
facilitate the identification of systems of strongly lensed images, where an image system consists of all the gravitationally-lensed counter-images from the same background galaxy.
The greater sensitivity of JWST to small flux fluctuations and improved spatial resolution 
enable searches for the fainter caustic or micro-caustic crossing events of stars at $z<2$ such as Icarus \citep{Kelly2018NatAsCCE}, or Warhol \citep{Chen2019CCE,Kaurov2019CCE}, stars between $2<z<5$ such as Godzilla \citep{Diego2022}, stars at $z\approx6$ such as Earendel \citep{welch22} or even further up to the first stars as proposed in \cite{Windhorst2018ApJSCCE}.\\

JWST is already providing breakthrough results in the study of the most distant galaxies. Pointing JWST towards gravitational lenses results in a more powerful combined telescope with an effective diameter a factor $\sqrt{|\mu|}$ times larger, where $|\mu|$ is the absolute magnification of the lensed background object. For galaxies, typical magnifications can reach factors of a few tens, making the combination JWST+SL cluster similar to a space telescope analogue to JWST but with with 20-30\,m diameter. In the context of high-z object detection, the lensed source counts are initially lower than a blank field due to the reduction in search volume from magnification. At the same time, the blank field luminosity function flattens with higher redshift, allowing the number of lensed galaxies to overtake the number of blank field galaxies \citep[{\it e.\,g.},][]{Mahler2019MACS0417,Salmon2020HighzRelics,Pascale2022a}.
\par
For much smaller background objects like magnified stars near caustics or in general for sub-pc structures near caustics, the underlying magnification factor can be of order $\sim1000$. This translates into effective apertures for JWST of $\approx 200$ meters! 
Even without the ultra-high magnification boosts near the critical lines, SL has allowed us to probe several magnitudes deeper than blank fields, {\it i.\,e.}, down to rest-frame UV luminosities $M_{\mathrm{UV}}\lesssim-13$ magnitudes \citep{bouwens17a,bouwens22a,bouwens22b,livermore17,atek18,ishigaki18, Vanzella2021} and stellar masses $M_{\star}\gtrsim10^6\,\mathrm{M}_{\odot}$ \citep{bhatawdekar19,kikuchihara20,furtak21,strait21}. With the redder wavelength range, greater sensitivity, and resolving power of the JWST, we can expect many more galaxies at even higher redshifts to be detected in the coming months. In order to fully characterize them and study their physics, it is beneficial to have different models which are constructed in independent works.  After this paper was submitted to the archive, we have become aware of two new lens models based on \texttt{Lenstool} which include both MUSE spectroscopic redshift constraints and JWST imaging  \citep{Mahler2022, Caminha2022}.
%we will need accurate SL magnification models.

%At least two lensing models using the \texttt{glafic} \citep{Oguri2010_25clusters} and \texttt{Lenstool} \citep{Jullo2007Lenstool} codes (see also, e.g. \citealt{FoxMahlerSharon2022}) have already been generated for SMACS0723 and are publicly available on the RELICS website\footnote{\url{https://archive.stsci.edu/prepds/relics/}}. In addition, another \texttt{Lenstool} model was recently constructed (J. Richard, private communication) using similar constraints to those used here, including the three spectroscopic redshifts from MUSE, and is publicly available online as well\footnote{\url{https://cral-perso.univ-lyon1.fr/labo/perso/johan.richard/ALCS_models/SMACS0723/}}. These models are considered \emph{parametric}, in the sense that the cluster galaxies and dark matter (DM) components are assumed to follow known profile shapes which can be modeled using analytic or parametrized formulae. Here, we present a new model with the \texttt{Light-Traces-Mass} approach \citep[\texttt{LTM};][]{Broadhurst2005a,Zitrin2009_cl0024,Zitrin2014CLASH25}. On account of being different in nature than parametric techniques, the \texttt{LTM} method allows to probe a different range of solutions. This will be very important for high-redshift studies in this cluster, for which a representative range of possible magnifications for background galaxies is needed. In addition, given the much deeper JWST data and their longer wavelength coverage, we expect that more lensed galaxies will be uncovered soon using these data, and so the model could be refined further and compared to the existing pre-JWST version.

In this paper we identify 14 new image families using the JWST data and present a new \emph{parametric} SL model. The paper is organized as follows: In \S \ref{s:data} we detail the NIRCam imaging and photometry needed to provide constraints for the lens model, which in turn is discussed in  \S \ref{s:code}.
%In \S \ref{s:data} we detail previous observations of the cluster which are relevant to constructing the lens model, which is described in \S \ref{s:code}.
In \S \ref{s:results} we present and discuss the results. The conclusion appears in \S \ref{s:summary}. Throughout this work we use a $\Lambda$CDM cosmology with $\Omega_{M}=0.3$, $\Omega_{\Lambda}=0.7$, and $H_{0}=70$ km~s$^{-1}$~Mpc$^{-1}$. Unless otherwise stated, we use AB magnitudes \citep{Oke1983ABandStandards}, and errors correspond to $1\sigma$.

\section{Observations and Data}\label{s:data}
%%%%%%%%%%%%%%%%%%%%%%%%%%%%%%%%%%%%%%%%%%%%%%%%%%%
Observations with the \textit{Near Infrared Camera} \citep[NIRCam; {\it e.\,g.},][]{rieke05} aboard the JWST were executed as part of the Director's Discretionary Time on 2022 June 06 (PI: Pontoppidan; Program ID 2736). Exposures were taken in F090W, F150W, and F200W in the short-wavelength (SW), and F356W, F277W, and F444W in the long-wavelength (LW) channels, totalling 12.5\,h of integration time. \textit{Near Infrared Spectrograph} \citep[NIRSpec;][]{jakobsen22} observations were made in the same field as a part of the same program. Two exposures were taken in each of the F170LP and F290LP filters, totalling 11787 sec of integration time in each filter. SMACS0723 was also observed using the \textit{Mid Infrared Instrument} \citep[MIRI;][]{bouchet12,rieke15} and the \textit{Near Infrared Imager and Slitless Spectrograph} \citep[NIRISS;][]{doyon12}. The JWST data analyzed in this work can be found on MAST:\dataset[10.17909/rwdx-k029]{http://dx.doi.org/10.17909/rwdx-k029}.

The raw data from all four instruments were processed using   the JWST science calibration pipeline which performed the background subtraction, flat-fielding, correction for cosmic ray hits, correction for image distortions, re-pixelization by the drizzle approach onto a common astrometric reference frame, co-addition to make the mosaic,  and combination into a  color image. For the NIRCam data, we reduce the raw data products independently from \textsc{uncal} files to \textsc{i2d} mosaics using a custom pipeline based on JWST pipeline version \textsc{1.6.2} alongside with the Calibration Reference Data System (CRDS) version \textsc{0942}. The pipeline includes updates based on on-flight calibrations that account for zero point offsets between the filters and the NIRCam modules \citep{Rigby2022}, critical for accurate photometry and derived properties. Furthermore, this pipeline includes additional steps over the default JWST pipeline to improve on its astrometry and background subtraction. For a detailed description on the additional steps see \cite{Adams2022}.

Using the NIRCam images based on our independent reduction with improved calibration, we extract photometry for our targets of interest using forced photometry of small, 0.32 arcsecond diameter circular apertures centered on the peak flux of the target in each band. The photometry is aperture corrected for a point spread function using the simulated point spread functions (PSFs) from \texttt{WebbPSF} \citep{Perrin2014}. This information is then fed into the photometric redshift code \texttt{LePhare} \citep{ArnoutsLPZ1999,Ilbert2006BPZ}. \texttt{LePhare} operates by fitting a large grid of galaxy spectral energy distributions to the photometry provided. We run \texttt{LePhare} with a suite of galaxy templates from \citep{BC03}, allowing for E(B-V) values between 0 and 1.5 \citep{Calzetti2000}, redshifts between 0 and 12 and applying the \citep{Madau1995} treatment for absorption from the IGM. The redshifts provided in Table \ref{multTable} are the redshift of peak probability and errors generated from a $\chi^2$ grid produced by \texttt{LePhare} based on the templates, redshifts and dust attenuation combinations available.  Several entries are missing values as a result of
various sources of contamination or by a lack of detection. We refer to the Appendix for more details. We note that these NIRCam data
%in the  SMACS0723 field 
can be compared with bluer data using HST taken as a part of the RELICS program (PI: D. Coe), and the \emph{Spitzer Space Telescope} (PI: M.~Bradac).  

We also made use of the available NIRSpec data set.  In particular, we analyzed those spectra which were acquired at the positions of our image systems  to confirm our image system designations. We refer to  \S 3.1 for more details. We also consulted the MIRI images to test for achromaticity within each of our image systems, for those systems that were detected. 

\begin{figure}[h]
    \centering
    \includegraphics[width=8.5cm]{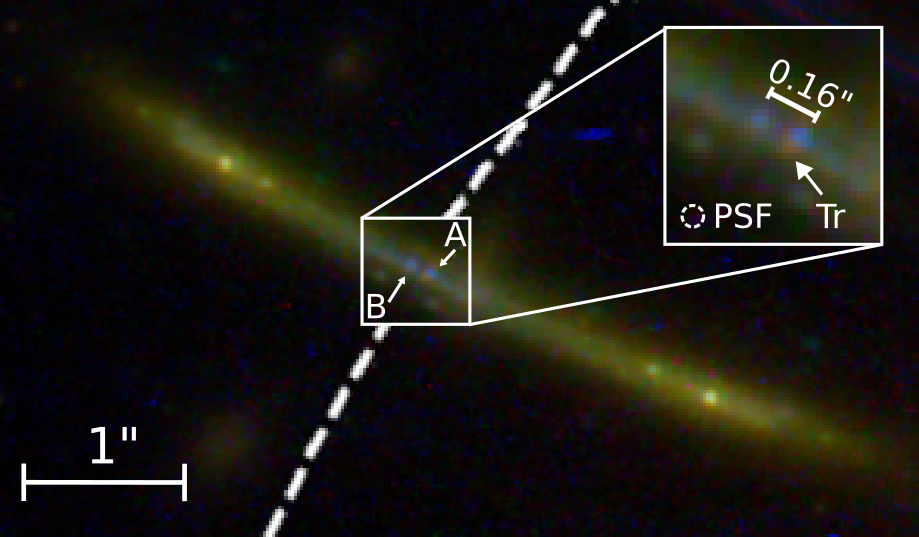}
    \caption{Close-up view of system 5 with the $z=1.43$ critical curve marked in the dashed line on this F090W+F150W+F200W composite color image. Two compact sources in the inset, knots A and B, have a separation of $\sim0.16^{\prime \prime}$, and have sizes that are  consistent with the width of the PSF. The knots bracket the critical curve that must pass in between them at an angular separation of $\sim0.08^{\prime \prime}$. We estimate high magnification factors as a result of this close proximity to the critical curve of 
    %Based on the proximity to the critical curve, we estimate a magnification factor of 
    $|\mu|\sim$750 for each source. A red knot is also detected at a position slightly offset from knot A, Tr, which may be a microlensed transient. We refer to \S \ref{s:results} for more information regarding this potential microlensed transient.}
    \label{fig:arc5}
\end{figure}

Overall, the arcs in this field make up a rich tapestry of distorted images typical of lensed sources and at a level of detail that has never been seen before. The increased spatial resolution  reveals lensing constraints even down to the small substructures within the arcs, 
%that can be used as individual lensing constraints, 
since many of these small scale sources are equally multiply-imaged, such as the double arc in system 5 (Figs.~\ref{fig:imstamps} and \ref{fig:arc5}). 
Meanwhile on the scale of the cluster, thanks
%Another surprise is that thanks 
to the remarkably dark infrared sky background of JWST images, we can get a high signal-to-noise view of the intra-cluster light (ICL).
This cluster has an ICL that is  elongated and nonuniform.  We refer to \S 4 for a discussion of how the ICL's relatively unique features suggest that this cluster is not dynamically-relaxed.

\begin{figure*}[t!]
    \centering\includegraphics[width=18cm]{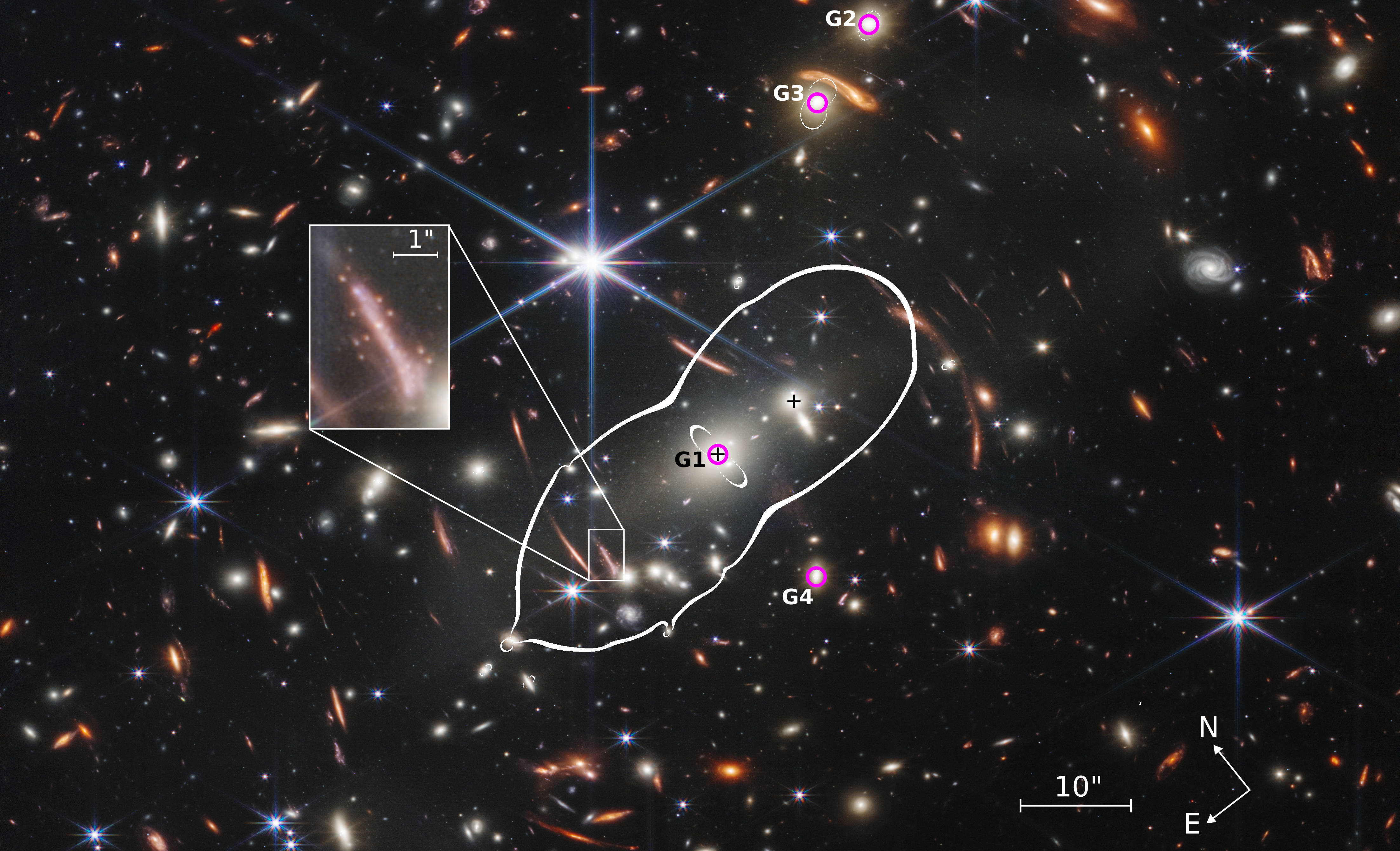}
    \caption{Critical curve at $z=1.45$ from the SMACS 0723 lens model is overlaid as a color composite image generated from the six existing JWST NIRCam filters depicting the central 1.5 by 1 arcminutes of the cluster. The model incorporates strong-lensing constraints from the 19 image families identified in total, 14 of which are discovered by this work using the JWST data). 
    The two ``+" symbols mark the positions of the two dark matter halos used in this model, and the 4 magenta rings mark the galaxies whose weight is left free in the model. 
    %Shown in the inset is 
    Image 2.2 is rare for containing $\geq10$ bright knots located at large projected galactocentric radii (inset).}
    \label{fig:curve}
\end{figure*}

\section{SL modeling of SMACS0723}\label{s:code}
We generate a new SL model exploiting the JWST data in which we identify the new sets of multiple images, as is described in \S\ref{sec:multiple-images}. Cluster member galaxies used in the modeling are described briefly in \S\ref{sec:cluster_members}. The modeling method is described in \S\ref{sec:SL-code}.

% Our \texttt{LTM} SL model of SMACS0723 takes two components as inputs: The strongly lensed multiple image systems described in section~\ref{sec:multiple-images} as SL constraints and the cluster member galaxies identified in section~\ref{sec:cluster_members} as one cluster mass component. We also give a brief summary of our modeling methods in section~\ref{sec:SL-code}.

\subsection{New multiple images in SMACS0723} \label{sec:multiple-images}
%------------------------------------------------------------------
We identify new lensing constraints taking advantage of the superior quality of the JWST images. We start by adopting the first five multiple image systems published in \citet{Golubchik2022}, following their numbering scheme. These systems serve also as a model-independent guide to search for  new systems.
For example, galaxies at close projected separations and at similar redshifts typically maintain similar spatial distributions when distorted by a gravitational lens. Moreover, amongst the galaxy images, counter images with negative and positive parities are in general symmetric to one another with respect to the critical curve. After visual inspection  of the new JWST/NIRCam image, and corroboration by available photometric and/or spectroscopy redshift constraints and model-predicted locations, we identify {\it 14} new sets of multiply lensed galaxies which are reported in Table~1.

Fortunately, there are spectroscopic redshifts available for some of the image systems. The redshifts of image systems 1, 2, and 5 are $z=1.450\pm0.001$, $z=1.378\pm0.001$ and $z=1.425\pm0.001$, respectively and were measured with MUSE \citep{Golubchik2022}. Note that while JWST/NIRSpec spectroscopic redshifts of 35 galaxies have been reported in this field with redshifts as high as 8.3 as reported by the JWST press  release,
an investigation of the NIRSpec spectroscopy uncovered spectra for only one of our 42 images from our 19 image families: image 4.1. However, an inspection of the spectrum in each of the spectral band-passes did not produce any significant features that were present in multiple visits and therefore did not yield a redshift. While valuable, further spectroscopic redshifts are however not crucially necessary for securing our image system identifications which relies on a large set of image multiplicities vetted in several ways.  We note that NIRISS spectroscopy of SMACS0723 is also available but not used in this study. Its redshift constraints will inform future lens models.

\subsection{Cluster member galaxies} \label{sec:cluster_members}
For our model we use the same the cluster member galaxy selection as in \citet{Golubchik2022}, which was based on the red sequence shown therein (their Fig.~4). In addition, we add here 3-4 galaxies not used in the initial model by \citet{Golubchik2022} which lie near the bright star 20$^{\prime\prime}$ north of the BCG (which hindered somewhat their previous detection), and are included as they may affect the reproduction of nearby counter images. A revision of this selection using spectroscopic data (such as {\it e.g.}, the NIRISS data), is deferred to future work.

\subsection{SL modeling method} \label{sec:SL-code}
%------------------------------------------------------------------
The \texttt{LTM} model presented in \citet{Golubchik2022} required a very strong external shear, indicating that the cluster's central mass distribution is highly elongated. Since the \texttt{LTM} approach is limited in the intrinsic matter ellipticity it can accommodate, and for comparison, we make use here of a new, fast parametric method that we recently constructed (A. Zitrin; in preparation; the method can also be referred to as analytic, {\it i.\,e.}, it is not limited to a grid's resolution). The method is similar in nature to other parametric lens modeling techniques such as \texttt{Lenstool} \citep{Jullo2007Lenstool}, \texttt{Glafic} \citep{Oguri2010_25clusters} or \texttt{GLEE} \citep{Halkola2006,Grillo2015_0416}. In such parametric methods, dark matter halos can be elliptical which more easily accounts for the required elongation (which is also evident from the distribution of arcs).

The models in our new parametric method have two main components. First, the cluster members galaxies are modeled with double pseudo elliptical mass distributions (dPIE; \citealt{Eliasdottir2007}) which are assumed spherically symmetric with the exception of the BCG, and are defined following the prescriptions of \cite{Jullo2007Lenstool} and \cite{Zitrin2013M0416}:
\begin{equation}
\begin{cases}
\sigma = \sigma^{*}(\frac{L}{L^{*}})^{1/4} \\
r_{cut} = r_{cut}^{*}(\frac{L}{L^{*}})^{\alpha} \\
r_{core} = r_{core}^{*}(\frac{L}{L^{*}})^{1/4}
\end{cases}
\end{equation}
Where the $\sigma$ is the velocity dispersion, $r_{cut}$ is the cut-off radius, $r_{core}$ is the core radius, and $L^{*}$ is the typical luminosity of a galaxy at the cluster redshift. We fix $r_{core}^{*}=0.2$ kpc for all dPIEs, use an $L^{*}$ equivalent to a galaxy of $m^{ref}_{F814W} = 22.13$, and assume a constant mass-to-light ratio ($\alpha=0.5$), while $\sigma^{*}$ and $r_{cut}^{*}$ are left free to be optimized by the model.

The second modeling component is the set of cluster DM halos. These can in principle be modeled either as elliptical Navarro-Frenk-White (NFW) profiles \citep{Navarro1996}, or as pseudo elliptical mass distributions (PIEMD; {\it e.\,g.}, \citealt{Monna2015Scaling}); or dPIEs. The third and last component which can be added if necessary is a two-component external shear.

The method is similar to our previous parametric implementation outlined in \citet{Zitrin2014CLASH25} which has been well vetted, but is not limited to the assigned grid resolution. The new improved version used here has been already applied to various clusters and has also been tested on simulated clusters, accommodating both image- and source-plane minimization. The minimization of the model is done via a Monte-Carlo Markov Chain (MCMC) with a Metropolis-Hastings algorithm \citep[{\it e.\,g.},][]{Hastings1970MCMC}. We include annealing in this procedure, and the chain typically runs for several dozen thousand steps after the burn-in stage. Errors are calculated from the same MCMC chain.

\subsection{SMACS0723 SL model}

For modeling SMACS0723 by this method, we constrain the model using 16 of the 19 identified image systems and a total of 48 images (2 sets of multiple images are used from system 5; see Table~\ref{multTable}), including spectroscopic redshifts for 3 systems and leaving all other system redshifts to be fit by the model. We represent cluster members as dPIE profiles, as explained above, and the cluster DM halos as PIEMDs ({\it i.\,e.}, the main free parameters for each halo aside from the ellipticity and its position, are a core radius and a velocity dispersion). We use two DM halos, centered on the two central galaxies respectively (marked by ``o" symbols in Fig.~\ref{fig:curve}). We do not incorporate an external shear. Since galaxies can deviate from the assumed scaling relations, we leave the four brightest galaxies to be freely weighted (marked by rings in Fig~\ref{fig:curve}), which means each of their velocity dispersions $\sigma$ is scaled by a weighting parameter optimized by the model (see Table \ref{lensTable}; G1-G4).
The BCG's ellipticity and position angle are also left free to be optimized by the model. In total, there are 29 free parameters in the model: the normalization for the galaxy velocity dispersion $\sigma^{*}$, the normalization for the galaxy cut-off radius $r_{cut}^{*}$, 4 free parameters for each of the 2 dark matter halo PIEMDs, 4 galaxy weights, the BCG ellipticity and position angle, and the 13 redshifts for systems without spectroscopic redshifts.
For the lensed sources, each multiply-imaged system should converge to a single position in the source plane.
To enforce this constraint, we minimize the differences between the image positions in the source plane following the prescription of \citet{Keeton2010GReGr..42.2151K},
which usually converges to a solution relatively quickly and without loss of quality, as shown by
\citet{Keeton2010GReGr..42.2151K} in their successful \textit{Hubble Frontier Fields} \citep[HFF;][]{Lotz2016HFF} lens models. Given the 48 lensed images, this constraint amounts to 62 SL constraints on the model (two for each lensed image with the exception of one image per multiplicity), which gives a total of 33 degrees of freedom for the model.

\section{Results and Discussion}\label{s:results}
%%%%%%%%%%%%%%%%%%%%%%%%%%%%%%%%%%%%%%%%%%%%%%%%%%%%

The best-fit SL model, which is the one for which $\chi^2$ is minimized, is presented in Fig.~\ref{fig:curve}. Critical lines are overlaid for $z_s$\,=\,1.45 ({\it i.\,e.}, the redshift of image system 1).
The figure illustrates the elongated configuration of SMACS0723. The model was minimized using a positional uncertainty of $\sigma_{pos}=0.5^{\prime \prime}$, resulting in a $\chi^{2}$ of 159.4 ($\chi^2_{\nu}=4.8$) in the image plane, and an \emph{rms} of 0.48\arcsec\, in reproducing the positions of the multiple images; the optimized model parameters can be found in Table~\ref{lensTable}.

\begin{figure}
    \centering
    \includegraphics[width=8.5cm]{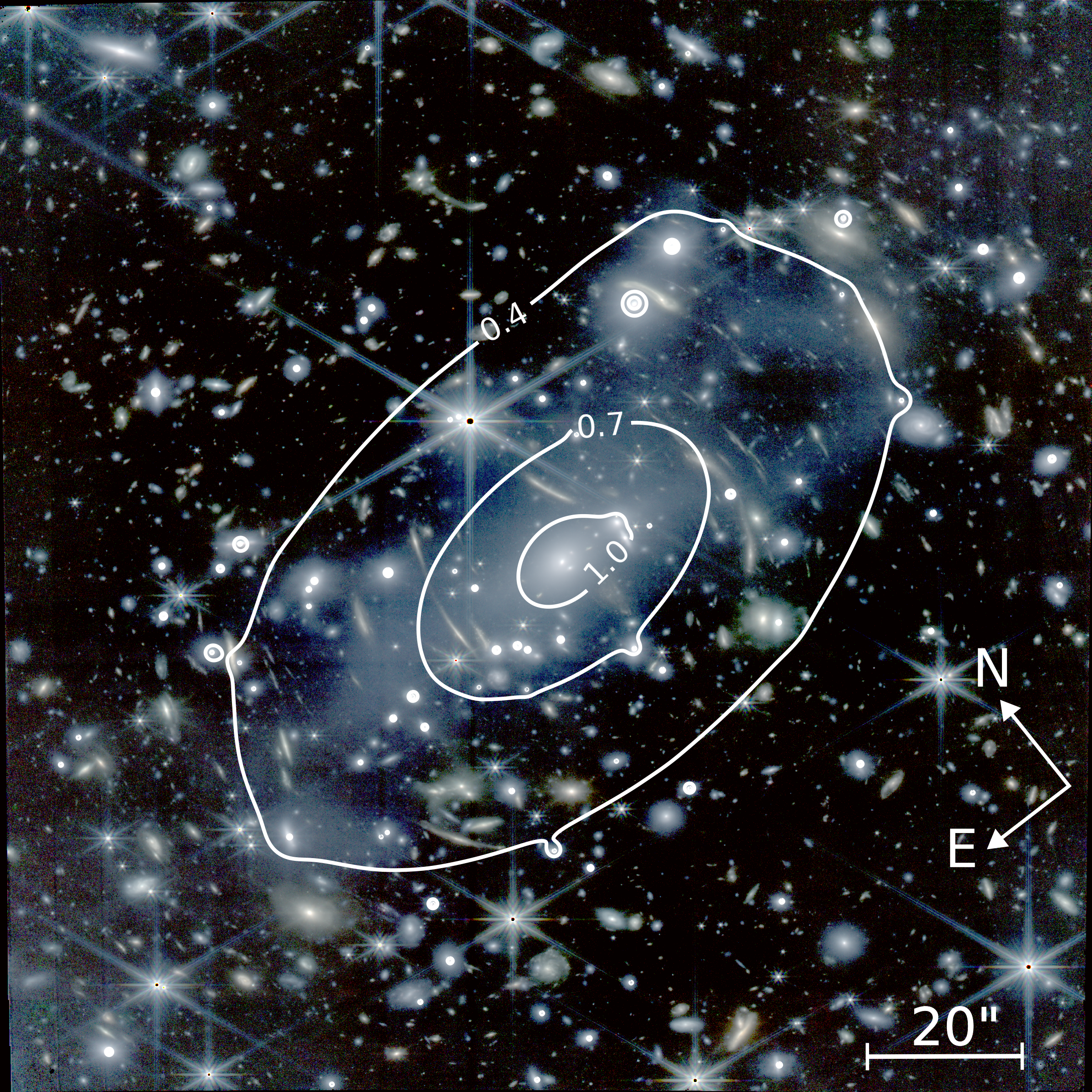}

    \caption{The intracluster light for SMACS 0723 in a F277W+F356W+F444W color composite image. The ICL structure is especially clear in F277W and F356W, and is much less prominent in the SW bands. The pronounced ICL extends along the long-axis of the cluster similar to the mass distribution from the lens model, as shown by the overlaid contours of projected mass density $\kappa$.  Typically a smoothly-varying feature, in SMACS0723 the ICL has significant structures, including a large ``loop" feature in the northwest component, and a large lobe-like feature in the southeast component.}
    \label{fig:ICL}
\end{figure}

The effective Einstein radius at a given redshift is 
computed as $\theta_E$\,=\,$\sqrt{A/\pi}$, with A being 
the area enclosed within the critical curves.  At 
$z$\,=\,2, $\theta_E$\,=\,18.4$^{\prime \prime}$\,$\pm$\,1.8$^{\prime \prime}$, 
and the mass contained inside the critical curve is
(5.91\,$\pm$\,0.83)\,$\times$10$^{13}$~M$_{\odot}$, which is overall consistent with the values of $\theta_E$\,=\,16.9$^{\prime \prime}$\,$\pm$\,2$^{\prime \prime}$, and $M$\,=\,$(4.15$\,$\pm$\,0.58)$\times$\,10$^{13}$ M$_{\odot}$ obtained in the HST \texttt{LTM} model \citep{Golubchik2022}. The uncertainties on the Einstein radius and mass are similar for each of the models and are limited by the systematics.

The JWST/NIRCam images uncover fine details of the faint substructures within the arcs.  In particular, two compact sources in system 5, knots A and B (5.A and 5.B in Table~\ref{multTable}), are spatially-resolved  on opposite sides of the critical curve, constraining the position of the critical curve in our lens model (Fig.~\ref{fig:arc5}). The image pair has a close angular separation to the critical curve of 0.08$^{\prime \prime}$, implying extremely high magnifications for each image and making it an ideal system in the search for caustic crossing events. Near to the critical curve, magnification follows an inverse relationship with distance $\mu = \mu_0/D_{crit}$, where $\mu_{0}$ is dependent on the slope of the lensing potential and can be measured from the lens model \citep{welch22,diego2019}. Uncertainty in $\mu_{0}$ is accounted for by sampling the model MCMC chain, while the uncertainty in $D_{crit}$ is assumed to be the uncertainty in the centroid position of the images. Our parametric model implies a magnification of $\mu = 737^{+1553}_{-454}$ for each image, whose positions are reproduced to within $0.02^{\prime \prime}$ by the model.
\par 
Magnification near the critical curve is highly sensitive to systematics, and may vary significantly between modeling approaches \citep{Meneghetti2017}. \cite{Meneghetti2017} demonstrated that the uncertainty in magnification can reach greater than 30\% at $\mu>10$ across all modeling approaches.
 Hence statistical errors in magnification from the model are most likely underestimate the true error which is dominated by systematics.
Assuming these images are unresolved, this magnification could imply parsec or even subparsec sizes for this object. Given the arc's close proximity to the cluster center, however, microlensing from intracluster stars could work to smooth out the critical curve, placing an upper limit on the persistent magnification possible \citep{Venumadhav2017,Dai2021}.

Consider again the image pair consisting of knots A and B in Fig.~\ref{fig:arc5}.  There is a red knot offset from knot A that is detected only in some of the NIRCam bands, called Tr (see appendix section \ref{sec:arc5}). We present the argument that Tr is neither knot A nor knot B.
Assuming knots A and B are counter images, the pair will always be detected together in any given band. Since the separation of these mirrored knots is about 0.16$^{\prime \prime}$, which is larger than the FWHM of any JWST filter in this dataset (F444W, FWHM = 0.145$^{\prime \prime}$, is the largest), JWST should be able to spatially resolve this pair. We detect knots A and B in all the SW filters, 
but detect only a single peak in each of the LW filters near to the position of knot A (see appendix, section \ref{sec:arc5} and Fig. \ref{fig:multiwav}).
Since the image pair is not detected together through the LW filters, this single peak must refer to Tr. Hence, Tr is {\it distinct} from knots A and B. 
The lack of a counterimage, and its placement in close proximity to the critical curve, suggest Tr is a microlensing event. However we cannot rule out that it is a red foreground galaxy. Given the very red colors and small angular sizes involved, confirmation of Tr as a transient would require JWST follow-up observations.
We note that there are no other similar dual-color detections in any other objects in system 5 that would suggest an image misalignment. 
\begin{figure}
    \centering
    \includegraphics[width=9.5cm,trim=2cm 0.25cm 2cm 1.5cm,clip]{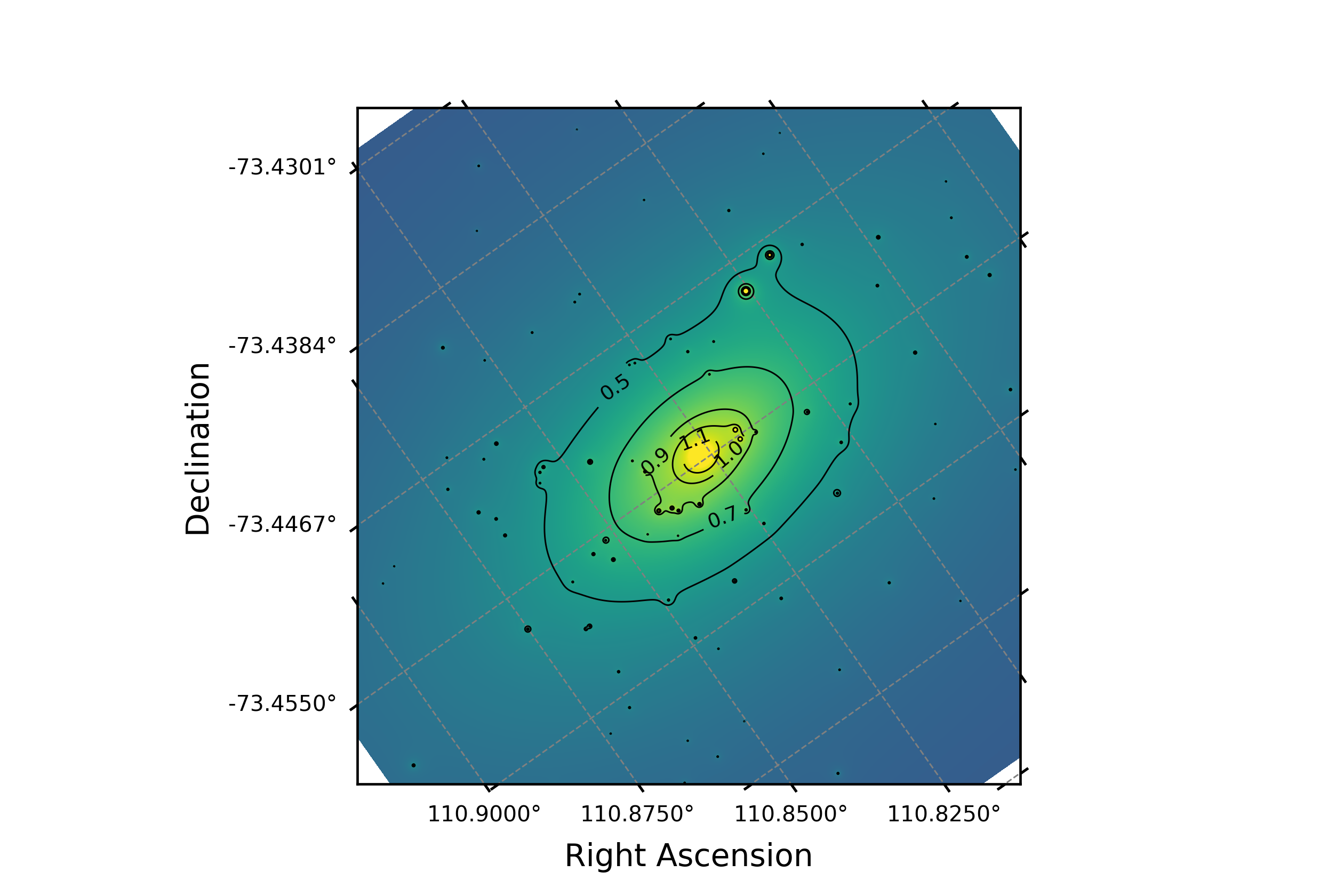}
    \caption{Projected mass density of SMACS0723. We show
$\kappa$, the surface mass density distribution in units of the critical density for strong lensing, for the source redshift of system 1, at $z=1.45$, plotted on a grid of constant RA and Dec (dashed lines). The resulting mass distribution exhibits a clear ellipticity along the same axis as the ICL as shown in Fig.~\ref{fig:ICL}}.
    \label{fig:kappa}
\end{figure}

The giant arcs also present some rather unusual features, such as system 2 with a spectroscopic redshift of $z$\,=\,1.38 first measured by \citet{Golubchik2022}.  Image 2.2 has $>$10 prominent clumps all of a similar red color and extending to high galactocentric radii (see Fig.~3, inset). 
While significant sub-kpc structure of star formation regions is not unusual for star-forming galaxies at $z=1-3$ (or even sub-100pc, {\it e.\,g.}, \citealt{Johnson:2017ab}), the very large separations cast some doubt that the extended knots are identical in nature to those in what might be the plane of a galactic disk. 
These clumps also appear to be different (redder) than those typically detected in clumpy galaxies  \citep[{\it e.\,g.},][]{Shibuya:2016aa}.
The physical characteristics are
more 
consistent with numerical predictions for globular clusters. For example, \cite{Sameie2022} found in hydrodynamic simulations that the longest lived globular clusters likely formed at high redshifts beyond the half-light radius (up to hundreds of parsecs away from their host galaxy).
\cite{Pozzetti2019} go on to predict the color
evolution 
for high redshift globular clusters, but also point out that NIRCam colors are most likely insensitive to the formation redshift.  \citet{Vanzella2017} have reported detections of candidate young ($<$10 Myr) globular clusters which are blue and at high redshifts of $z$\,=\,3--6. 
Based on this information, it is tempting to speculate that the clumps seen in system 2 may be redder and older ($\sim$Gyr) counterparts of those young candidate globular clusters detected at the much higher redshifts.
 Based on SED-fitting from $0.4-4.4\mu$m, \cite{Mowla2022} found evidence that the clumps were consistent with evolved globular clusters with ages of $3.9-4.1$ Gyr, implying formation a mere $\sim$0.5 Gyr after the Big Bang. We note that 
another possible explanation,
is that the
red clumps fitting the description of those seen in system 2 may arise as a result of ram pressure stripping of dwarf galaxies in the dense environments near in projection to its disk \citep[{\it e.\,g.},][]{Mayer2006,Boselli2022}. 
Further follow-up, such as detailed source plane reconstruction and photometric SED modeling, is needed to uncover the nature of these sources.

A striking feature of this new JWST/NIRCam image is the morphology of the baryonic component responsible for the intracluster light (ICL), which exhibits a loop-like feature in the northwest component and a large lobe in the southeast component. We perform our own reduction of the raw images to mitigate the variations in sky background across chips as described in \S \ref{s:data}, although some
background noise can also be seen in each of the four corners of the image due to unoptimized sky subtraction in the JWST reduction pipeline (Fig.~\ref{fig:ICL}). The ICL is due to stars stripped away from their galaxies but still gravitationally bound to the cluster. Earlier work has suggested that the ICL is a good tracer of the DM distribution (or vice-versa) since both stars in the ICL and DM are expected to behave as collisionless particles and thus respond only to gravity \citep{Montes2018b}. 
Our initial SL model exhibits a similar macro-structure to the ICL as seen in Fig.~\ref{fig:ICL}. The ICL also extends 
beyond the current set of SL constraints, but in combination with weak lensing derived from the JWST images it should be possible to produce a lens model that covers the entire range of the ICL in a way that gives insights as to the intriguing apparent correlation between the ICL and the DM distribution. In particular, the presence of these large-scale features in the ICL may motivate detailed simulations to examine the merging scenarios that constrain the time since the major merger and estimate the relative velocities involved. 

We note that after this paper was submitted to the archive, and prior to submission to ApJ, we have become aware of two other papers on SL modeling in this field \citep{Mahler2022, Caminha2022}. There are some differences in the results presented in these other papers, but we have checked and found that including their results into our lens model does not alter the results of this study.

\section{Summary}\label{s:summary}
We presented a new JWST parametric model for the massive cluster SMACS0723, incorporating new multiple image family constraints identified in the JWST press-release color image and in the NIRCam data set. Our model  builds upon the existing lens models \citep{Golubchik2022} by increasing the SL constraints with 14 newly identified image families, which represents a factor of three improvement over the HST model. 
This parametric model has a large impact because it provides a starting point for modeling SL clusters in the era of JWST. The mass map reveals a somewhat complex and extended mass profile and, in the detection of the image pair knots A and B opposite the critical curve in system 5, also opens a route to the study of caustic transients
in the discovery of 
an apparent microlensed transient adjacent to knot A. The ICL has a shape that is broadly-similar to that of the galaxies and the model, but with rare larger-scale features that appear to retain the memory of its dynamical history, a question which can be investigated further with numerical simulations.
In the future, spectroscopic and photometric redshifts will bolster the reliability of our lens model, allowing also a more precise placement of the critical curve needed to engage in caustic transient studies. 

\bigskip
\bigskip

\section*{acknowledgements}
We wish to thank the anonymous referee, whose suggestions have improved this manuscript.
We would also like to thank the RELICS and PEARLS collaborations for data products or discussions that have stimulated this work, and of course the people behind the JWST, its instruments and public release of this incredibly beautiful first data set.

MP was funded through the NSF Graduate Research Fellowship grant No.~DGE
1752814, and acknowledges the support of System76 for providing computer
equipment. BLF thanks Ori Ganor and the Berkeley Center for Theoretical Physics for their hospitality during the writing of this paper. The BGU group acknowledges support by Grant No. 2020750 from the United States-Israel Binational Science Foundation (BSF) and Grant No.~2109066 from the United States National Science Foundation (NSF), and by the Ministry of Science \& Technology, Israel. CC and NA acknowledge support from the European Research Council (ERC) Advanced Investigator Grant EPOCHS (788113). J.M.D. acknowledges the support of project PGC2018-101814-B-100 (MCIU/AEI/MINECO/FEDER, UE) Ministerio de Ciencia, Investigaci\'on y Universidades and support by the Agencia Estatal de Investigaci\'on, Unidad de Excelencia Mar\'ia de Maeztu, ref. MDM-2017-0765. LD acknowledges the research grant support from the Alfred P.
Sloan Foundation (Award Number FG-2021-16495).

This work is based on observations made with the NASA/ESA \textit{Hubble Space Telescope} (HST) and NASA/ESA/CSA \textit{James Webb Space Telescope} (JWST) obtained from the \texttt{Mikulski Archive for Space Telescopes} (\texttt{MAST}) at the \textit{Space Telescope Science Institute} (STScI), which is operated by the Association of Universities for Research in Astronomy, Inc., under NASA contract NAS 5-03127 for JWST, and NAS 5–26555 for HST. These observations are associated with program 14096 for HST, and 2736 for JWST. This work is also based on observations made with ESO Telescopes at the La Silla Paranal Observatory obtained from the ESO Science Archive Facility. The authors thank the Centre for Astronomy and Particle Theory (CAPT) of University of Nottingham for providing all computational infrastructure to run the JWST Reduction pipeline, and Philip Parry for technical support. LF acknowledges financial support from CAPES - Brazil. We also thank Anthony Holloway and Sotirios Sanidas at JBCA for critical help with computer infrastructure that made this work possible.

This research made use of \texttt{Astropy},\footnote{\url{http://www.astropy.org}} a community-developed core Python package for Astronomy \citep{astropy13,astropy18} as well as the packages \texttt{NumPy} \citep{vanderwalt11}, \texttt{SciPy} \citep{virtanen20}, \texttt{matplotlib} \citep{hunter07}, and some of the astronomy \texttt{MATLAB} packages \citep{maat14}. 

\newpage

\newpage

\appendix

\section{Arc systems in SMACS0723}
We present the list of all multiply-imaged systems below.  The image families are vetted in multiple ways:  by their similar morphological components, similar colors, similar redshifts (photometric and/or spectroscopic), and/or consistency with the lens model. The IDs are given in the first column, where the ``?" indicates an image member candidate, by which we mean that it fails one or more of the above criteria, or there is more than one candidate that fits these criteria. 
The photometric redshift estimates, $z_{NIRCam}$, are computed  using the SED fitting code \texttt{LePhare}, as discussed in the main text, with matched-aperture photometry of the six bands of NIRCam imaging as described in \cite{Adams2022}. An entry has no $z_{NIRCam}$ value if  the lensed source was significantly contaminated by a stellar diffraction spike, a bright cluster member galaxy or other projection effects, or if it was too faint to be detected in two or more of the NIRCam bands.
The model-predicted redshifts, $z_{model}$ and $z_{NIRCam}$ are in relative agreement with the exception of five systems. The discrepancies in systems 3, 4, 12, and 15 can be attributed to a degeneracy in the SED fitting for z=6-7 and z=1.5-2.5, which comes as a result of the Lyman Break and Balmer Break both falling between F090W and F150W at these redshifts respectively. The inclusion of F115W could break this degeneracy in most cases, and may be necessary for any future surveys of the redshift distribution of background galaxies. The other outlier is image 16.1, for which $z_{NIRCam}-z_{model}>3\sigma$. This system is faint and elongated, and is most likely not well represented by the 0.32$^{\prime \prime}$ circular aperture imposed on these data, in addition to being contaminated by the ICL.
We also note that that the lens models are limited by the arc systematics.  Lens predictions may change as image system constraints become improve, for example by the measurement of new spectroscopic redshifts.

\section{Image Pair in System 5}\label{sec:arc5}
\begin{figure*}[t!]
    \centering\includegraphics[scale =0.2]{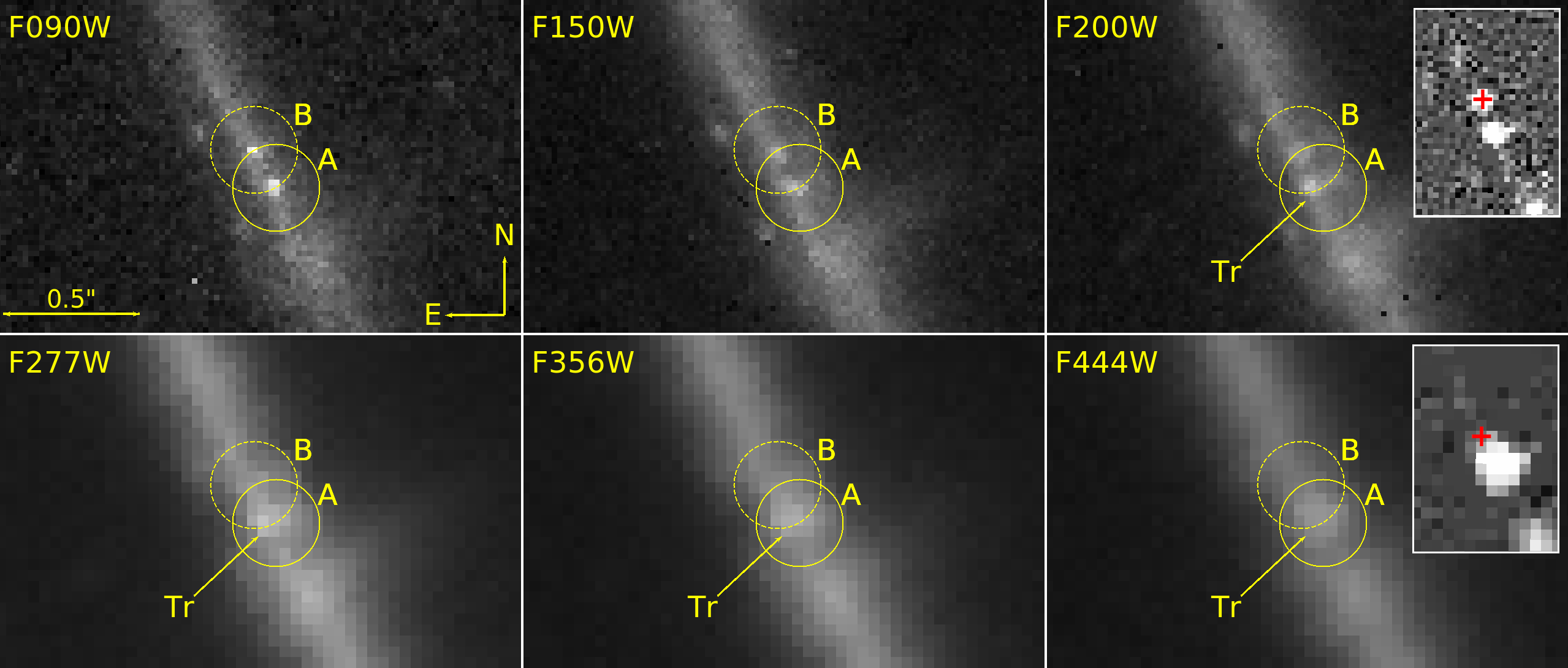}
    \caption{The image pair A and B of system 5 across all six available JWST filters. In each filter we center on the positions of A and B with rings of radius $0.16^{\prime \prime}$, which is roughly equal to their angular separation. The F090W, F150W, F200W filters exhibit two separate knots consistent with the marked positions of A and B. F200W, however, also shows the bright source Tr that is slightly offset from the position of knot A. 
    Only a single peak is detected in each of the LW filters, which is 
    also slightly offset from the position of knot A. The insets depict the F200W and F444W images with the diffuse arc background subtracted off and the position of knot B indicated by a ``+" sign. Since knot A and knot B are counterimages of one another and must appear together, the Tr object seen in F200W and the LW filters is instead understood to be a candidate microlensing event.}
    \label{fig:multiwav}
\end{figure*}

Knots A and B of system 5 are counterimages which are situated opposite the critical curve and thus likely highly magnified. As shown in Fig. \ref{fig:multiwav}, 
the image pair appears in each of the SW filters and is spatially resolved, with an angular separation of $\sim0.16^{\prime \prime}$. In SW filter F200W, the microlensed transient, Tr, is detected as well; it is bright and offset from the position of knot A, resulting in the red knot seen in Fig. \ref{fig:arc5}. In the LW filters, the knots A and B drop out, leaving only the redder Tr visible as a single peak.

\begin{deluxetable*}{ccccc}
%\begin{longtable){ccccc}
\tablecaption{List of arc systems}
\label{multTable}
\tablecolumns{5}
%%\tablewidth{0.85\linewidth}
\tablehead{ \colhead{ID$^a$} & \colhead{R.A}     & \colhead{Decl.} & \colhead{$z_{NIRCam}^b$} & \colhead{$z_{\mathrm{model}}$} \\
                            & \colhead{J2000.0} & \colhead{J2000.0}  &   \colhead{}  & \colhead{} }
%\endfirsthead
%\endhead
\startdata
1.1 & 07:23:21.8012 & -73:27:03.467 & 1.45$^c$ & \\
1.2 & 07:23:22.3430 & -73:27:17.077 & 1.45$^c$ & \\
1.3 & 07:23:21.4174 & -73:27:31.329 & 1.45$^c$ & \\
\hline
2.1 & 07:23:21.3288 & -73:27:03.369 & 1.38$^c$ & \\
2.2 & 07:23:21.8238 & -73:27:18.337 & 1.38$^c$ & \\
2.3 & 07:23:20.8184 & -73:27:31.378 & 1.38$^c$ & \\
\hline
3.1$^e$ & 07:23:19.3406 & -73:26:54.554 & -- & 1.81 [1.65 -- 1.91]\\
3.2$^e$ & 07:23:19.6709 & -73:27:18.562 & -- & "\\
3.3$^e$ & 07:23:18.0861 & -73:27:34.507 & $6.65^{+0.05}_{-0.06}$ & "\\
3.4$^e$ & 07:23:17.6180 & -73:27:17.238 & -- & "\\
\hline
4.1 & 07:23:13.3027 & -73:27:16.373 & $6.64^{+0.03}_{-0.05}$ & 2.02 [1.94 -- 2.48]\\
4.2 & 07:23:13.7402 & -73:27:30.116 & $2.16^{+4.54}_{-0.67}$ & "\\
4.3 & 07:23:15.2214 & -73:26:55.308 & $2.20^{+4.46}_{-0.17}$ & "\\
\hline
5.1 & 07:23:17.7775 & -73:27:06.454 & 1.43$^c$ & \\
5.2 & 07:23:17.4012 & -73:27:09.692 & 1.43$^c$ & \\
5.3 & 07:23:17.0777 & -73:27:36.422 & 1.43$^c$ & \\
5.A & 07:23:17.5937 & -73:27:08.213 & 1.43$^c$ & \\
5.B & 07:23:17.6067 & -73:27:08.073 & 1.43$^c$ & \\
\hline
\hline
!6.1 & 07:23:20.6382 & -73:27:06.170 & $2.66^{+0.04}_{-1.48}$ & 1.44 [1.27 -- 2.02]\\
!6.2 & 07:23:20.8642 & -73:27:10.805 & $1.37^{+1.29}_{-0.23}$  & "\\
6.3 & 07:23:20.2857 & -73:27:39.168 & -- & "\\
\hline
7.1 & 07:23:10.9765 & -73:26:54.832 & -- & 6.27  [2.95 -- 8.38]\\
7.2 & 07:23:10.7520 & -73:26:56.676 & -- & "\\
7.3 & 07:23:11.9016 & -73:26:49.389 & $3.66^{+0.04}_{-0.91}$ & "\\
\hline
8.1 & 07:23:12.6279 & -73:27:36.552 & -- & 5.93  ($>$ 4.67)\\  
8.2 & 07:23:11.9399 & -73:27:19.101 & -- & "\\ 
8.3 & 07:23:15.1738 & -73:26:48.223 & $6.26^{+0.18}_{-0.1}$ & "\\   
\hline
9.1 & 07:23:12.7404 & -73:27:17.641 & $2.69^{+0.11}_{-0.03}$ & 2.90 [2.57 -- 3.14]\\
9.2 & 07:23:13.2719 & -73:27:32.036 & -- & "\\
9.3 & 07:23:15.0263 & -73:26:53.194 & $1.72^{+0.28}_{-0.04}$ & "\\
\hline
10.1$^d$ & 07:23:17.6727 & -73:27:06.073 & $1.22^{+1.42}_{-0.05}$ & \\   
10.2$^d$ & 07:23:17.2165 & -73:27:09.979 & $1.26^{+1.41}_{-0.06}$ & \\    
10.3$^d$ & 07:23:16.9616 & -73:27:36.214 & $1.19^{+0.09}_{-0.10}$ & \\   
\hline
11.1 & 07:23:14.4676 & -73:27:21.001& -- & 1.66 [1.22 --1.98]\\
11.2 & 07:23:14.6783 & -73:27:25.882 & -- & "\\
\hline
12.1 & 07:23:17.3278 & -73:26:56.724 & -- &  1.56 [1.52 -- 1.71]\\
12.2 & 07:23:15.5468 & -73:27:15.540 & $6.75^{+0.05}_{-0.05}$ & "\\
12.3 & 07:23:16.2054 & -73:27:33.241 & $6.79^{+0.03}_{-0.02}$ & "\\
\hline
13.1 & 07:23:19.1576 & -73:26:56.027 & $1.73^{+1.03}_{-0.03}$ & 2.34 [2.29 -- 2.77]\\
13.2 & 07:23:17.3314 & -73:27:14.825 & -- & "\\
13.3 & 07:23:17.6167 & -73:27:41.842 & $2.69^{+0.05}_{-1.02}$ & "\\
13.4 & 07:23:19.8353 & -73:27:15.697 & -- & "\\
\hline
14.1$^d$ & 07:23:12.4242 & -73:27:29.857 & -- & 1.83 [1.68 --  2.03]\\
14.2$^d$ & 07:23:12.4965 & -73:27:31.974 & -- & "\\
14.3$^d$ & 07:23:12.5821 & -73:27:32.465 & -- &" \\
\hline
15.1? & 07:23:16.7116 & -73:26:55.135 & -- & 2.21 [1.67 --2.29]\\
15.2 & 07:23:14.7745 & -73:27:16.356 & $6.59^{+0.05}_{-0.07}$ & "\\
15.3 & 07:23:15.4182 & -73:27:32.298 & $2.30^{+0.06}_{-0.20}$ & "\\
\hline
\enddata
\end{deluxetable*}

% BBEAK TABLE HERE SO IT APPEARS IN TWO PAGES
\setcounter{table}{0}
\begin{deluxetable*}{ccccc}
%\begin{longtable){ccccc}
\tablecaption{Cont.}
\label{multTable}
\tablecolumns{5}
%\tablewidth{0.85\linewidth}
\tablehead{
\colhead{ID$^a$
} &
\colhead{R.A
} &
\colhead{DEC.
} &
\colhead{$z_{NIRCam}$
} &
\colhead{$z_{\mathrm{model}}$
} \\  
 &J2000.0&J2000.0&&  
}
\startdata
\hline
\hline
16.1 & 07:23:16.9967 & -73:27:09.328 & $0.29^{+0.06}_{-0.07}$ & 1.07 [0.82 --1.04]\\
16.2 & 07:23:16.9296 & -73:27:10.256 & -- & "\\
\hline
17.1  & 07:23:17.5975 & -73:27:20.571 & $2.34^{+0.31}_{-0.10}$ & 1.72 [1.68 -- 1.92]\\
17.2  & 07:23:17.7931 & -73:27:26.849 & -- & "\\
17.3 & 07:23:19.1779 & -73:26:50.524 & -- & "\\
\hline
18.1 & 07:23:17.2313 & -73:27:02.060 & -- & 1.33 [1.24 -- 1.37]\\
18.2 & 07:23:16.0440 & -73:27:13.482 & -- & "\\ 
18.3 & 07:23:16.3551 & -73:27:32.183 & -- & "\\
\hline
GL.1$^{d}$  & 07:23:21.8176 & -73:27:41.814 & -- &1.31 [1.15 -- 1.58]\\
GL.2$^{d}$  & 07:23:22.4453 & -73:27:40.964 & $1.49^{+0.88}_{-0.13}$ & "\\
GL.3?$^{d}$ & 07:23:22.4781 & -73:27:41.727 & $1.49^{+0.88}_{-0.13}$ & "\\
\hline
Einstein cross & 07:23:02.7975 & -73:27:08.814 & & \\
\enddata
\tablecomments{\emph{Column~1:} ID; \emph{Columns~2 \& 3:} Right Ascension and Declination; \emph{Column~4:} Redshift. For systems 1, 2 and 5 we quote the spectroscopic redshift from MUSE \citep{Golubchik2022}, and note that the NIRISS spectrum for system 1 confirms the stated redshift. 
\emph{Column~5:} The redshift of the system as predicted from the SL model. Candidate images whose identification is not secure are marked with ``?." Note that these were therefore not used in the minimization.}
\tablenotetext{a}{{The first five image systems are drawn from \citet{Golubchik2022}.}}
\tablenotetext{b}{These are the photometric redshift estimates using the six bands of NIRCam imaging. 
}
\tablenotetext{c}{These redshifts are spectroscopic values from \citet{Golubchik2022}.}
\tablenotetext{d}{These systems were not used to constrain the parametric SL model, but were confirmed in the WSLAP+ model.}
\tablenotetext{e}{\cite{Mahler2022} find a spectroscopic redshift of 1.99 for this system using MUSE, but this is not included in the model.}
\end{deluxetable*}

\newpage

\begin{deluxetable*}{ccccc}
\tablecaption{SL Model Optimized Parameters}
\label{lensTable}
\tablecolumns{5}
\tablehead{\colhead{} & \colhead{$e$} & \colhead{$\theta [^{\circ}]$} & \colhead{$\sigma[kms^{-1}]$} & \colhead{$r_{cut}[kpc]$}}
\startdata
Main DM Halo & $0.46^{+0.07}_{-0.05}$ &  $0.12^{+0.02}_{-0.02}$ & $1276.50^{+60.28}_{-76.43}$ & $84.43^{+7.32}_{-6.90}$ \\
West DM Halo & $0.36^{+0.33}_{-0.25}$ & $0.29^{+0.26}_{-0.16}$ & $52.98^{+41.18}_{-41.04}$ & $14.91^{+10.25}_{-10.55}$\\
BCG & $0.16^{+0.10}_{-0.09}$ & $17.09^{+2.05}_{-2.41}$ & --- & ---\\
\hline
\hline
Scalings$^{a}$ & $N_{gal}=133$ & $m^{ref}_{F814W} = 22.13$ & $\sigma^{*} = 101.44^{+15.08}_{-16.10}$ & $r_{cut}^{*} = 48.45^{+7.05}_{-6.37}$ \\
\hline
\hline
Weights$_{gal}^{b}$ & $G1=1.87^{+0.10}_{-0.11}$ & $G2=1.34^{+0.48}_{-0.31}$ & $G3=1.64^{+0.27}_{-0.29}$ & $G4=0.91^{+0.11}_{-0.15}$
\enddata
\tablecomments{\emph{Column~1:} Object; \emph{Column~2:} Ellipticity; \emph{Column~3:} Position Angle; \emph{Column~4:} Velocity Dispersion. \emph{Column~4:} Cutoff Radius. All errors are given as the 68.3\% Confidence Interval.}
\tablenotetext{a}{The scaling relations for all galaxies.}
\tablenotetext{b}{Weights for the four galaxies allowed to deviate from the scaling relations, as labeled in Fig.~\ref{fig:curve}.}
\end{deluxetable*}

\bibliographystyle{aasjournal}
\bibliography{bibfile}

\end{document}